\documentclass[conference]{IEEEtran}
\usepackage{amsmath,graphicx}
\usepackage{amssymb}
\usepackage{psfrag}
\usepackage{lipsum}
\usepackage{hyphenat}
\usepackage{cite}

\usepackage[monochrome]{color}

\def\Htran{\mbox{\tiny $\mathrm{H}$}}
\def\Ttran{\mbox{\tiny $\mathrm{T}$}}
\def\CN{\mathcal{N}_{\mathbb{C}}}
\def\Real{\mathbb{R}}
\def\Complex{\mathbb{C}}
\def\Integer{\mathbb{Z}}

\def\Ex{\mathbb{E}}
\def\sinc{\mathrm{sinc}}
\def\rank{\mathrm{rank}} 

\def\imagunit{\mathsf{j}} 

\usepackage[dvipsnames,svgnames]{xcolor}
\usepackage[textwidth=30mm]{todonotes}
\newcommand{\vect}[1]{{\bf{#1}}}

\def\BibTeX{{\rm B\kern-.05em{\sc i\kern-.025em b}\kern-.08em
    T\kern-.1667em\lower.7ex\hbox{E}\kern-.125emX}}

\title{\vspace{-0.0cm}Degrees of Freedom of \\Holographic MIMO Channels\vspace{-0.3cm}}
\author{\IEEEauthorblockN{Andrea Pizzo and Thomas L. Marzetta}
\IEEEauthorblockA{\textit{Department of Electrical and Computer Engineering} \\
\textit{New York University, USA}\\
\texttt{\{andrea.pizzo,tlm\}@nyu.edu}}\vspace{-0.9cm}
\and
\IEEEauthorblockN{Luca Sanguinetti}
\IEEEauthorblockA{\textit{Dipartimento di Ingegneria dell'Informazione} \\
\textit{University of Pisa, Italy}\\
\texttt{luca.sanguinetti@unipi.it}}\vspace{-0.9cm}
}

\begin{document}
\maketitle
\begin{abstract}
We consider spatially-constrained apertures of rectangular symmetry and aim to retrieve the limit to the {\color{red}average} number of spatial degrees of freedom (DoF), obtained elsewhere through different analyses and tools. Unlike prior works, we use the Fourier plane-wave series {\color{red}expansion}, recently introduced in \cite{pizzo2019spatial}, where a statistical model for the small-scale fading in the far-field is developed on the basis of a continuous-space and physics-based orthonormal expansion over the Cartesian spatial Fourier basis.
This expansion yields a set of statistically independent random coefficients whose cardinality directly gives the limit to the {\color{red}average} number of DoF.
The treatment is limited to an isotropic scattering environment but can be extended to the non-isotropic case through the linear-system theoretic interpretation of plane-wave propagations.
\end{abstract}
\vspace{0.2cm}
\begin{IEEEkeywords}
Spatial degrees of freedom, Fourier theory, physics-based channel modeling, Cartesian coordinates.
\end{IEEEkeywords}


\vspace{-0.0cm}
\section{Introduction}
\label{sec:intro}
\vspace{-0.0cm}

A \emph{holographic MIMO} (multiple-input multiple-output) array consists of a massive (possibly infinite) number of antennas into a compact space \cite{pizzo2019spatial,Marzetta2018,Bjornson2019d}.
This concept is also known as large intelligent surfaces \cite{Rusek2018,Renzo2019a}, and holographic beamforming \cite{Black2017}.
 
In MIMO systems, the capacity grows linearly with the number $\eta$ of spatial degrees of freedom (DoF) \cite{TseBook,Telatar,Foschini}, which is determined by the scattering environment and antenna array geometries at the transmitter and receiver sides \cite{TseBook}. 
{\color{red}In its asymptotic form, a Holographic MIMO system can be thought of as the ultimate form of a spatially-constrained MIMO and consists of a transmit and receive array taking the form of spatially-continuous electromagnetic apertures, where the number of antennas $N$ at both sides goes to infinity \cite{Marzetta2018,pizzo2019spatial}.
Therefore, $\eta$ is limited by the scattering environment and the resolution at which these apertures can resolve it \cite{Foschini}.
A fundamental question thus arises: given an area limitation on the apertures, what are the average number $\eta$ of spatial DoF of a spatially-continuous holographic MIMO system?}
To answer this question, continuous-space channel models have been used in the literature under different propagation conditions and aperture geometries, e.g. \cite{Poon,Franceschetti,Rusek2018} among others.
These models are physics-based and thus driven by electromagnetic theory considerations.
In \cite{Poon}, the authors use a physics-based and continuous-space channel model to compute the number of DoF of apertures of spherical symmetry (e.g., segment, disk, sphere) under a \emph{deterministic} monochromatic scattering channel. A signal space approach is used that is based on an orthonormal expansion of the channel over polar and spherical spatial Fourier bases. For a spherical aperture, $\eta$ is fundamentally limited by its {surface area} (rather than its volume), measured in units of wavelength-squared. This means that increasing $N$ does not increase $\eta$ indefinitely. This is analogous to a bandlimited waveform (time-domain) channel, given the bandwidth constraint $B$ and transmission interval $T$, increasing the number of time samples will not increase the capacity indefinitely. The available DoF are fundamentally limited to $2BT$ \cite{Shannon,GallagerBook}. The extension to a more general non-monochromatic environment and apertures of arbitrary geometry is provided in \cite{Franceschetti} and it is based on the celebrated Landau's eigenvalue theorem \cite{Landau1975}.

In \cite{Rusek2018}, the authors consider apertures of rectangular symmetry (e.g., segment, rectangle, parallelepiped) under line-of-sight (i.e., no scattering) propagation. This leads to a \emph{deterministic} channel, which is first used to evaluate the capacity, normalized by the aperture area, and then to compute the available DoF. In agreement with \cite{Poon,Franceschetti}, it turns out that the DoF per m of a segment deployment are fundamentally limited to ${2}/{\lambda}$
with $\lambda$ being the wavelength. For a rectangular deployment, the DoF per m$^2$ are limited to ${\pi}/{\lambda^2}$.

\begin{figure*}[t!]\vspace{-0.0cm}
\begin{equation}  \label{Fourier_planewave}\tag{5}
{h}_\pm(x,y,z) = \frac{1}{4 \pi \sqrt{\pi}}  \iint_{-\infty}^\infty  \sqrt{S_h(k_x,k_y)} W^\pm(k_x,k_y) e^{\imagunit \left(k_x x + k_y y \pm \gamma (\kappa_x,\kappa_y) z\right)} \,dk_x dk_y
\end{equation} 
\hrule\vspace{-0.5cm}
\end{figure*}

Like \cite{Rusek2018}, we consider apertures of rectangular symmetry but focus on a spatially-stationary \emph{random} monochromatic scattering propagation channel, which is {statistically} characterized by using the methodology developed in \cite{pizzo2019spatial}. Particularly, a signal space approach, that relies directly on an orthonormal expansion over the Cartesian spatial Fourier basis \cite{pizzo2019spatial}, is used to retrieve the limit to the number of DoF.
{\color{red}This expansion yields a set of statistically independent random coefficients from which the cardinality of this set directly gives the available DoF on average.}
We consider a rich scattering environment (i.e., isotropic propagation) particularly because it provides the upper limit to the number of DoF of any other scattering environment. The non-isotropic case can also be handled along the lines of \cite{Poon} still by using the general statistical model in \cite{pizzo2019spatial}. Also, we focus on the receiver side only since the transmitter can be treated similarly and the total number of spatial DoF will be given by the minimum of the two, as in classical MIMO systems.

As mentioned above, the basic maximum number of DoF result is a well-known result  \cite{Kildal2017} that can be proved by using different methodologies and approaches. However, unlike previous works, our paper obtains the same result through a new methodology \cite{Marzetta2018,pizzo2019spatial}; that is, a novel Fourier plane-wave series expansion of the random channel in Cartesian coordinates. To the best of our knowledge, there are no papers in the literature that look at the DoF in a generalized stochastic (not deterministic, which is the common way but it is unfortunately specific to the considered scenario only) spatially-stationary monochromatic channel.

\vspace{-0.0cm}
\section{Review of the Fourier-based Statistical Representation of Small-Scale Fading} \label{sec:preliminaries}
\vspace{-0.0cm}
We begin by reviewing the basics of the methodology developed in \cite{Marzetta2018,pizzo2019spatial}. 
Consider electromagnetic waves propagation in every direction (i.e., isotropic propagation\footnote{The extension to the non-isotropic case is briefly discussed in the conclusions and left for the extended version.}) through a homogeneous, isotropic, and infinite random scattered medium. 
Under these settings, electromagnetic waves without polarization qualitatively behave as acoustic waves \cite{Marzetta2018} and the 3D small-scale fading in the far-field can be generally modeled as a \emph{space-frequency scalar random field}
\begin{equation} \label{h_omega}
\left\{h_\omega(x,y,z):(x,y,z)\in\Real^3, \omega\in(-\infty,\infty)\right\} 
\end{equation}
which is a function of frequency $\omega$ and spatial Cartesian coordinates $(x,y,z)$.
We treat only monochromatic waves, i.e., propagating at the same frequency $\omega$, which can thus be omitted. We assume that $h(x,y,z)$ can be modeled as a zero-mean, spatially-stationary and Gaussian random field. The spatial autocorrelation function $
c_h (x,y,z) =  \Ex\{{h^*(x^\prime,y^\prime,z^\prime)} {h(x+x^\prime,y+y^\prime,z+z^\prime)}\}$ provides a complete statistical description of $h(x,y,z)$ in the spatial domain.
Alternatively, $h(x,y,z)$ can  be statistically described in the \emph{wavenumber}domain by its power spectral density \cite{Marzetta2018,pizzo2019spatial}:
\begin{align} \label{eq:PSD}
\!\!\!S_h(k_x,k_y,k_z)\! = \!\!\iiint_{-\infty}^{\infty}\!\! \!\!\!\!\! c_h(x,y,z) e^{-\imagunit (k_x x + k_y y + k_z z)} dx dy dz.\!\!\!
\end{align}


\subsection{Fourier Plane-Wave Spectral Representation}

The electromagnetic nature of the small-scale fading requires each realization of $h(x,y,z)$ to satisfy (with probability $1$) the scalar Helmholtz equation in the frequency domain, which reads (in a source-free environment) as $\left(\nabla^2 + \kappa^2  \right) h(x,y,z)  =  0 $ where $\kappa = {2 \pi}/{\lambda}$ is the wavenumber with $\lambda$ being the wavelength \cite[Eq.~1.2.17]{ChewBook}. As a direct consequence of the Helmholtz equation, we have that \cite{Marzetta2018,pizzo2019spatial}
\begin{equation} \label{PSD}  
S_h(k_x,k_y,k_z) = \frac{4\pi^2}{\kappa} \, \delta(k_x^2 + k_y^2 + k_z^2 - \kappa^2)
\end{equation}
which is an impulsive function with wavenumber support on the surface of a sphere of radius $\kappa$. This result is exploited in \cite{pizzo2019spatial} to represent $h(x,y,z)$ as the sum of two random fields: 
\begin{equation} \label{small_scale}
h(x,y,z) = h_+(x,y,z) + h_-(x,y,z)
\end{equation}
that are defined in \eqref{Fourier_planewave} where $W^\pm(k_x,k_y)$ are two 2D independent, zero-mean, complex-valued, white-noise Gaussian random fields and
\setcounter{equation}{5}
\begin{equation} \label{spectrum_iso}
S_{h}(k_x,k_y) = \frac{\pi}{\kappa}\frac{1}{ \gamma(k_x,k_y)}
\end{equation}
with 
\begin{equation} \label{eq:kappa_z}
\gamma(k_x,k_y) = \sqrt{\kappa^2 - k_x^2 - k_y^2}
\end{equation} 
is the 2D power spectral density of $h_\pm(x,y,z=0)$, defined over a compact support $k_x^2 + k_y^2 \le \kappa^2$ given by a disk $\mathcal{D}(\kappa)$ of radius $\kappa$ centered on the origin; see Fig.~2(b) in \cite{pizzo2019spatial}. The representation\footnote{Since $S_{h}(k_x,k_y)$ is singularly-integrable over $\mathcal{D}(\kappa)$ (i.e., $h(x,y,z)$ is a second-order field), the convergence of \eqref{Fourier_planewave} hold in the mean square sense \cite{pizzo2019spatial}.} in \eqref{Fourier_planewave} is called \emph{Fourier plane-wave spectral representation} \cite{pizzo2019spatial}. The ``plane-wave'' terminology refers to the fact that $h_\pm(x,y,z)$ represent a decomposition of the random field in terms of an \emph{uncountably infinite} number of plane-waves that are spatially propagating (known as \emph{propagating waves}) {\color{red}respectively through the left (upgoing) or right (downgoing) half-spaces created by an infinite plane passing through $(x, y, z)$ and perpendicular to the arbitrarily chosen $z-$axis.}
These waves have statistically-independent Gaussian-distributed random amplitudes $\sqrt{S_{h}(k_x,k_y)} W^{\pm}(k_x,k_y)$. 

Notice that the condition $k_x^2 + k_y^2 \le \kappa^2$ excludes a purely imaginary $\gamma(k_x,k_y)$ in \eqref{eq:kappa_z} that would account for the so-called \emph{evanescent waves} \cite{pizzo2019spatial}. These are excluded because decaying exponentially fast in space and thus not contributing to far-field propagation. This limits the bandwidth of $h(x,y,z)$ (in the wavenumber domain) to $|\mathcal{D}(\kappa)| = \pi \kappa^2$. This is a direct consequence of the Helmholtz equation and it is exploited in \cite{pizzo2019spatial} to obtain the following Fourier plane-wave series expansion, from which the DoF will be directly computed in Section III.





\vspace{-0.0cm}
\subsection{Fourier Plane-Wave Series Expansion}

Assume that the small-scale fading is observed over a continuous-space rectangular volume $(x,y,z)\in\mathcal{V} \subset \Real^3$ of finite side lengths $L_x,L_y$ and $L_z< \max(L_x,L_y)$ along the three spatial Cartesian axes.
In this case, only a \emph{countably finite} number of plane-waves can be resolved \cite{pizzo2019spatial}. These are obtained by uniformly sampling the wavenumber spectrum with spacing ${2\pi}/{L_x}$ and ${2\pi}/{L_y}$.
By rescaling the $k_x$ and $k_y-$axes as $\frac{L_x}{2\pi} k_x$ and $\frac{L_y}{2\pi} k_y$, the 2D lattice ellipse shown in Fig.~\ref{fig:disk_lattice} is obtained
\begin{equation}
\mathcal{E} = \{(\ell,m)\in\Integer^2 : \left({\ell \lambda}/{L_x}\right)^2 + \left({m \lambda}/{L_y}\right)^2 \le 1\}.
\end{equation}
The representation in \eqref{Fourier_planewave} can thus be approximated by a \emph{2D Fourier plane-wave series expansion} \cite[Eq.~(41)]{pizzo2019spatial}:
\begin{equation} \label{Fourier_series}
h(x,y,z) \approx  \mathop{\sum\sum}_{\ell,m\in \mathcal{E}} H_{\ell m}(z) e^{\imagunit 2\pi \left(\frac{\ell  x}{L_x} + \frac{m  y}{L_y}\right)}
\end{equation}
where the Fourier coefficients are the random variables
\begin{align} \label{Fourier_coeff_z}
H_{\ell m}(z) & = H_{\ell m}^+  e^{\imagunit \gamma_{\ell m} \; z} 
+  H_{\ell m}^- e^{-\imagunit \gamma_{\ell m} \; z} 
\end{align} 
with 
\begin{align} \label{gamma_ellm}
\gamma_{\ell m}= \gamma\left({2\pi\ell}/{L_x},{2\pi m}/{L_y}\right) 
\end{align} 
and $H_{\ell m}^\pm \sim \CN(0, \sigma_{\ell m}^2)$ being statistically-independent and circularly-symmetric Gaussian random variables with variances $\sigma_{\ell m}^2$ as computed in \cite[App.~D]{pizzo2019spatial}.
{\color{red}Given a pair $(\ell,m)$, $H_{\ell m}^\pm$ are associated respectively with the left or right half-spaces created by the infinite $z-$plane passing through $(x, y, z)$.}
Notice that \eqref{Fourier_series}  generates a periodic random field that approximates the small-scale fading $h(x,y,z)$ over its 2D fundamental period $L_x$ and $L_y$ (along the $x-$ and $y-$axes, respectively).
{The approximation error of \eqref{Fourier_series} reduces as $\min(L_x,L_y)/{\lambda}$ becomes large \cite{pizzo2019spatial}, and vanishes as ${\min(L_x,L_y)/{\lambda}\to\infty}$.  It follows that the approximation \eqref{Fourier_series} does not require the array size to be ``physically large'', but rather its normalized size (with respect to the wavelength $\lambda$). In practice, one may use the model at carrier frequencies sufficiently high while keeping the array size as constant in order to reach the desired accuracy.
At a carrier frequency ${f = 3}$~GHz (i.e., ${\lambda=10}$~cm), aperture lengths of ${L_x=L_y=1}$~m already provide ${\min(L_x,L_y)/{\lambda}=10}$, which increases to ${\min(L_x,L_y)/{\lambda}=10^2}$ at ${f = 30}$~GHz (i.e., ${\lambda=1}$~cm). Numerical results will show that these values are enough to achieve a good approximation. 

Assume now that the small-scale fading is observed over a line segment. Without loss of generality, assume that is observed along the $x-$axis with $x\in\mathcal{V} \subset \Real$ of length $L_x$. The Fourier plane-wave series expansion \eqref{Fourier_series} reduces to \cite[Eq.~(45)]{pizzo2019spatial}
\begin{equation}  \label{Fourier_series_linear}
h(x) \approx  \sum_{\ell=-L_x/\lambda}^{L_x/\lambda-1} H_\ell \, e^{\imagunit \frac{2\pi \ell x}{L_x}} 
\end{equation}
where $H_{\ell}\sim \CN(0, \sigma_\ell^2)$ with $\sigma_\ell^2$ given in \cite[App.~D]{pizzo2019spatial}.
The wavenumber support of $h(x)$ is limited to $k_x\in[-\kappa,\kappa]$ and can be obtained by collapsing the 2D lattice disk in Fig.~\ref{fig:disk_lattice} over the $k_x-$axis; that is, ${\kappa}$ represents the wavenumber bandwidth.

\vspace{-0.0cm}
\section{Degrees of Freedom}
\vspace{-0.0cm}
We now use \eqref{Fourier_series} and \eqref{Fourier_series_linear} to compute the number of DoF over linear ($\mathcal{V}_1$), planar ($\mathcal{V}_2$) and volumetric ($\mathcal{V}_3$) aperture spaces. 
 Before this, we review the reasoning behind the $2BT$ formula for waveform channels \cite{Shannon,GallagerBook}.

\begin{figure}[t!]
    \centering
     \includegraphics[width=0.9\columnwidth]{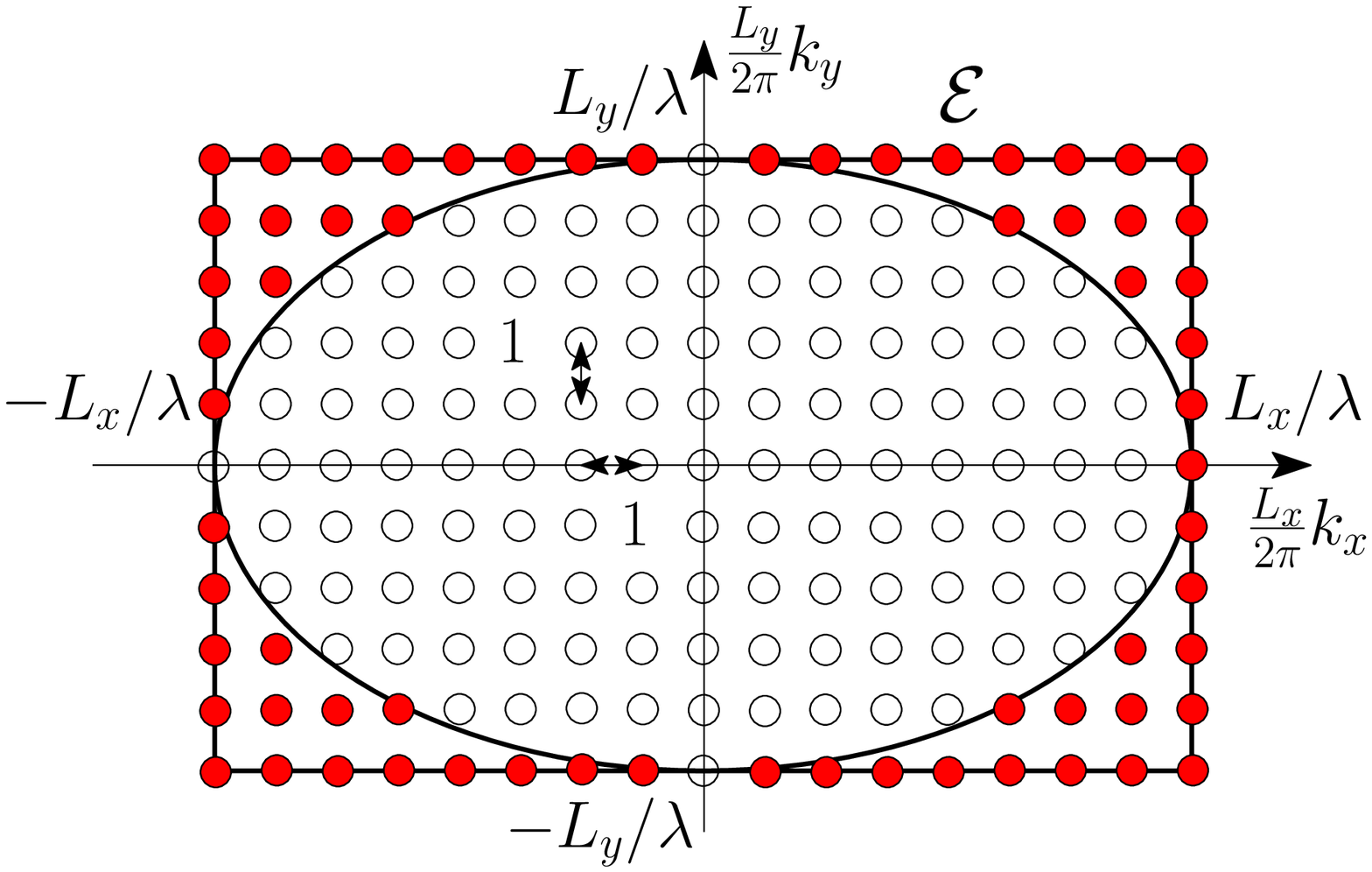} \vspace{-0.0cm}
  \caption{The 2D lattice ellipse $\mathcal{E}$ wavenumber spectral support of $h(x,y,z)$.}
   \label{fig:disk_lattice}\vspace{-0.5cm}
\end{figure}

Consider a bandlimited waveform channel $h(t)$ of bandwidth $B$ and time interval $[0,T]$. The Shannon-Nyquist sampling theorem states that we can approximate $h(t)$ as a linear combination of a countably finite number of elements of the cardinal basis of functions with coefficients collected inside the time interval and equally spaced by $1/2B$:
\begin{equation}\label{eq:Shannon}
h(t) \approx \sum_{n=-BT}^{BT-1} h\left(\frac{n}{2B}\right)\sinc\left(2B t - n\right), \quad t\in[0,T]
\end{equation}
where the approximation error becomes negligible as $B T\to\infty$.
The limit can be seen as $T$ going to infinity while $B$ going to zero, but $T$ has a higher convergence speed such that $B T\to\infty$. This is because physics-based signals are of limited energy and thus subject to the phenomena of \emph{spectral concentration} \cite{FranceschettiBook} under which as $T$ increases the effective bandwidth $B$ gets smaller and smaller. As a consequence, the available DoF are limited to a finite non-zero value 
\begin{equation} \label{DoF_time}
\eta = 2BT = \frac{\Omega}{\pi} T
\end{equation}
which is the product between time interval duration $T$ and the frequency bandwidth ${B=\Omega/2\pi}$.
Notably, \eqref{Fourier_series} and \eqref{eq:Shannon} are two bandlimited orthonormal series expansion having a countably-finite number of coefficients, whose cardinality determines the space dimension, i.e., the available DoF.

\vspace{-0.0cm}
\subsection{Linear arrays}
Assume that $h(x,y,z)$ is observed over a 1D line segment $\mathcal{V}_1$. Without loss of generality, assume that $\mathcal{V}_1$ is of length $L_x$ along the $x-$axis. From \eqref{Fourier_series_linear}, the Fourier plane-wave series expansion of $h(x) = h(x,0,0)$ is
\begin{equation}  \label{Fourier_series_1D}
h(x) \approx  \sum_{\ell=-L_x/\lambda}^{L_x/\lambda-1}   c_{\ell} \; \varphi_\ell(x) \qquad x\in \mathcal{V}_1
\end{equation}
where $\varphi_\ell(x) = \frac{1}{\sqrt{L_x}} e^{\imagunit \frac{2\pi \ell x}{L_x}}$
is the 1D Fourier basis and $c_{\ell} = \sqrt{L_x} H_{\ell}$.
Due to the statistical independence of the Fourier coefficients $H_{\ell}$, the average number of DoF is thus limited to {\color{red}the cardinality of $\{H_{\ell}\}$.} This yields (e.g., \cite{Rusek2018}):
\begin{equation}  \label{DoF_1D}
\eta_1 =\frac{2}{\lambda}L_x = \frac{\kappa}{\pi}L_x
\end{equation}
which states that the DoF are given by the ratio between the aperture length $L_x$ and half-wavelength, or equivalently, by the product between the aperture length $L_x$ and the wavenumber ${\kappa}$.
By comparing \eqref{DoF_1D} to \eqref{DoF_time}, we notice that there is a one-to-one correspondence between the two DoF formulas. In particular, \eqref{DoF_1D} is the spatial-wavenumber counterpart of \eqref{DoF_time} where the time interval $T$ and angular frequency bandwidth $\Omega$ are replaced by $L_x$ and $\kappa$, respectively.

\vspace{-0.0cm}
\subsection{Planar arrays}
Assume that $h(x,y,z)$ is observed over a 2D rectangle $\mathcal{V}_2$ of side lengths $L_x>L_y$.  From \eqref{Fourier_series}, the Fourier plane-wave series expansion of $h(x,y) = h(x,y,0)$ is
\begin{equation}  \label{Fourier_series_2D}
h(x,y) \approx  \mathop{\sum\sum}_{(\ell,m)\in \mathcal{E}} c_{\ell,m} \; \varphi_{\ell,m}(x,y) \qquad (x,y)\in \mathcal{V}_2
\end{equation}
where $\varphi_{\ell m}(x,y) = \varphi_{\ell}(x) \varphi_{m}(y)$ is the 2D Fourier basis, and $c_{\ell,m} = \sqrt{L_x L_y} H_{\ell m}(0)$. 
The average number of DoF is limited by the cardinality of $\{H_{\ell m}\}$, which is given by the measure of the wavenumber discrete support.
This is computed by counting the number of lattice points falling into the 2D lattice ellipse as reported in Fig~\ref{fig:disk_lattice}; that is, its Lebesgue measure $|\mathcal{E}|$.
By rewriting $\mathcal{E} = \{(\ell,m)\in\Integer^2 : Q(\ell,m)\le \alpha\}$ with $Q(\ell,m) = ({L_y}/{\lambda})^2 \ell^2 + ({L_x}/{\lambda})^2 m^2$ and $\alpha = \big({L_x L_y}/{\lambda^2}\big)^2$, the Lebesgue measure of $\mathcal{E}$ is $|\mathcal{E}| \approx 2\pi/\sqrt{D}$ with $D = 4 \alpha$ \cite[Eq.~(1.2)]{Nowak}. This yields
\begin{equation}  \label{DoF_2D}
\eta_2 = \frac{\pi}{\lambda^2} L_x L_y 
\end{equation}
which states that the DoF are proportional to the surface area of the aperture $\mathcal{V}_2$, measured in units of wavelength-squared. From \eqref{DoF_1D}, one may expect that the expansion of the field from a 1D segment into a 2D should yield $4L_xL_y/\lambda^2$ DoF. However, this is not the case. The DoF are reduced by a factor $\pi/4<1$, which is exactly the ratio between the areas of the disk $D(\kappa)$ and the square $R(\kappa)$ circumscribing it; see Fig.~2(b) in \cite{pizzo2019spatial}. This is due to the fact that evanescent waves do not contribute to far-field propagation.
Since this comes directly from the Helmholtz equation, which acts as a 2D linear space-invariant physical filter, we can think of $\pi/4 <1$ as a sort of physical filter inefficiency. 
Notably, this inefficiency is always $\pi/4$ regardless of the dimension of the 2D rectangular aperture.

\vspace{-0.0cm}
\subsection{Volumetric arrays}
Assume that $h(x,y,z)$ is observed over a 3D parallelepiped $\mathcal{V}_3$ of side lengths $L_x,L_y$ and $L_z< \min(L_x,L_y)$.  From \eqref{Fourier_series}, the Fourier plane-wave series expansion of $h(x,y,z)$ for any fixed $z$ is 
\begin{equation}  \label{Fourier_series_3D}
{h}(x,y,z) \approx  \mathop{\sum\sum}_{(\ell,m)\in \mathcal{E}} {c}_{\ell,m}(z) \; \varphi_{\ell,m}(x,y) \quad (x,y,z)\in \mathcal{V}_3\!\!
\end{equation}
where 
\begin{equation} 
c_{\ell,m}(z) = c_{\ell,m}^+ e^{\imagunit \gamma_{\ell m} z} + c_{\ell,m}^- e^{- \imagunit \gamma_{\ell m} z} 
\end{equation}
with $c_{\ell m}^\pm = \sqrt{L_x L_y} H_{\ell m}^\pm$.
From \eqref{Fourier_coeff_z}, we notice that ${h}(x,y,z)$ depends on $z$ only through the two complex exponentials $e^{\pm \imagunit \gamma_{\ell m} z}$ with $\gamma_{\ell m}$ given in \eqref{eq:kappa_z}. For every pair $(\ell,m)$ and fixed $z$, these two functions are completely known and do not carry any information.
Hence, we expect that the number of DoF over a 3D volume does not scale proportionally to $L_z$, as if it would have happened for $L_x$ and $L_y$. To show this, it is convenient to work with a single random vector \cite{VanTreesBook}.
We start by collecting an arbitrary number $N_z$ of samples along $z\in[0,L_z]$, i.e., $\vect{z}=[z_1,\ldots,z_{N_z}]^{\Ttran}$. This yields the 2D random vector field $\vect{h}(x,y) = h(x,y,\vect{z})$ given by
\begin{equation}  \label{Fourier_series_3D_vec}
\vect{h}(x,y) \approx  \mathop{\sum\sum}_{(\ell,m)\in \mathcal{E}} \vect{c}_{\ell,m} \; \varphi_{\ell,m}(x,y)
\end{equation}
where $\vect{c}_{\ell m} = \vect{A}_{\ell m} \; \bar{\vect{c}}_{\ell m} \in \Complex^{N_z\times 2}$ is a random vector of statistically-independent elements with
\begin{equation}   \label{A_3D}
\vect{A}_{\ell m} = \left[e^{\imagunit \gamma_{\ell m} \vect{z}}, e^{-\imagunit \gamma_{\ell m} \vect{z}} \right] \in  \Complex^{N_z \times 2}
\end{equation}
and $\bar{\vect{c}}_{\ell m} = [c_{\ell m}^+, c_{\ell m}^-]^{\Ttran} \in \Complex^{2\times 1}$.
Due to the statistical independence of the elements of $\vect{h}(x,y)$, the number of DoF are obtained by the product between \eqref{DoF_2D} and the space dimension spanned by $\vect{c}_{\ell m}$.
The latter is given by $\rank(\vect{A}_{\ell m})=2$, since the two columns of $\vect{A}_{\ell m}$ are linearly independent regardless of the choice of $\vect{z}$.
Thus, the DoF are limited to
\begin{equation}  \label{DoF_3D}
\eta_3 = \frac{2\pi}{\lambda^2} L_x L_y  
\end{equation}
which does not depend on $L_z$ (as intuitively anticipated).
Hence, the expansion of a 2D rectangular aperture into a 3D volumetric aperture asymptotically yields only a two-fold increase in the average number of DoF. The entire small-scale fading $h(x,y,z)$ is captured by two parallel planar surfaces.
Notice that if only one of the two half-spaces created by the aperture is contributing to $h(x,y,z)$, then $\eta_3 = \eta_2$ {\color{red}(e.g., \cite{Rusek2018})}.

\vspace{-0.0cm}
\section{Numerical results}
\vspace{-0.0cm}

\begin{figure}\vspace{-0.2cm}
     \centering
     \includegraphics[width=\columnwidth]{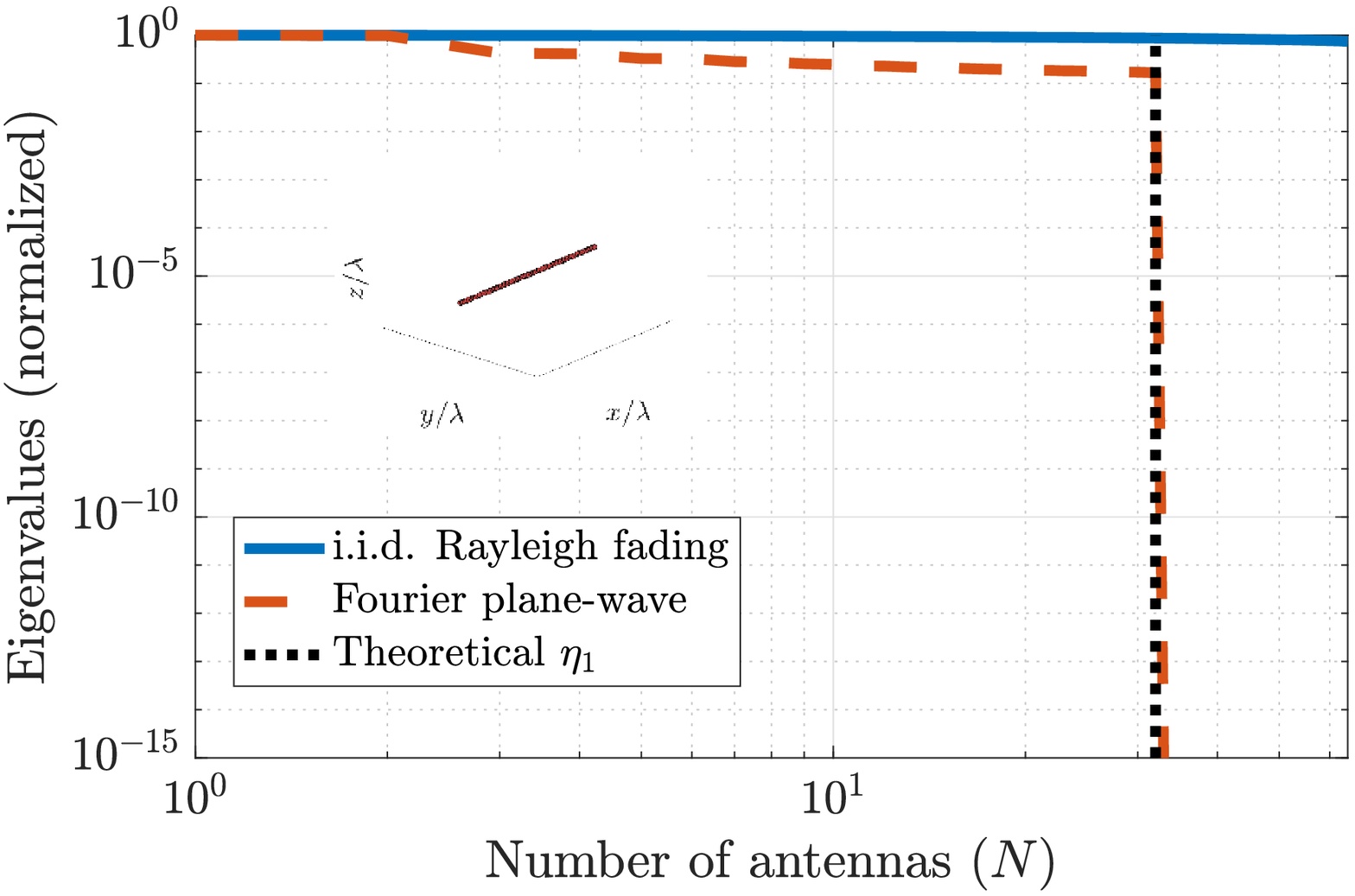} \vspace{-0cm}
  \caption{Eigenvalues of $\vect{H} \vect{H}^{\Htran}$ over $\mathcal{V}_1$ in descending order.\vspace{-0.0cm}} 
   \label{fig:1D_DoF}
\end{figure}

\begin{figure}\vspace{-0.2cm}
     \centering
     \includegraphics[width=\columnwidth]{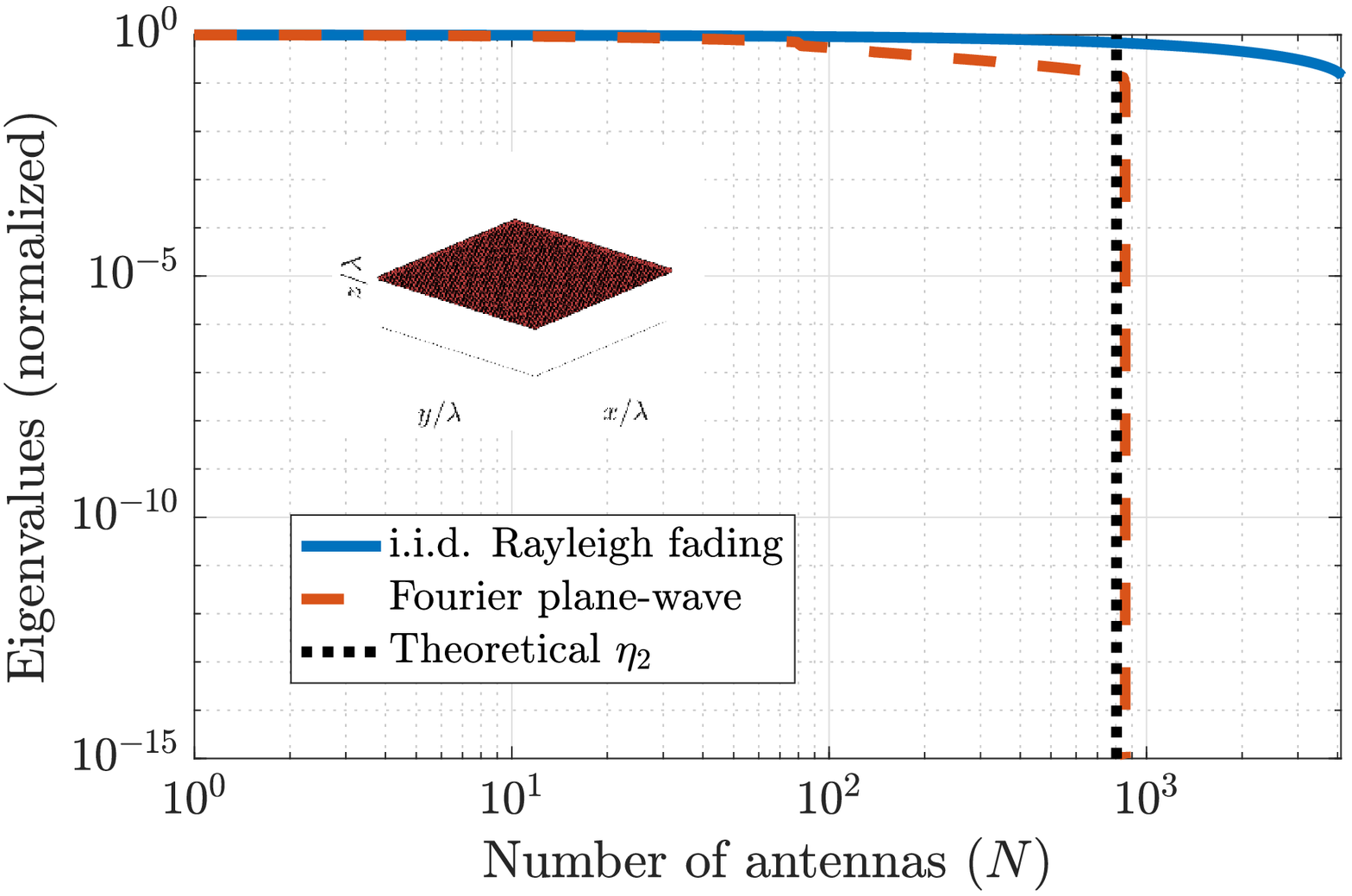} \vspace{-0cm}
  \caption{Eigenvalues of $\vect{H} \vect{H}^{\Htran}$ over $\mathcal{V}_2$ in descending order.\vspace{-0.5cm}} 
   \label{fig:2D_DoF}
\end{figure}

Numerical results are now used to validate the accuracy of our theoretical results for rectangular spaces $\mathcal{V}_i$ with $i=1, 2, 3$ of relatively small size compared to the wavelength $\lambda$. We numerically generate the continuous aperture by discretization on a spatial grid $\mathcal{V}_{N} = \{(x,y,z)_i \in \mathcal{V} : i=1,\ldots, N\}$ of $N$ points with uniform spacing $\Delta$ along all axes. We arbitrarily choose $\Delta = \lambda/4$. The channel samples generated by sampling \eqref{Fourier_series_1D}, \eqref{Fourier_series_2D}, and \eqref{Fourier_series_3D_vec} for any $(x,y,z)_i\in \mathcal{V}_{N}$ are collected into $\vect{h} \in\Complex^{N} $. 
To evaluate the space dimension spanned by the random vector $\vect{h}$ (and so the number of DoF) we first generate an ensemble $M$ of random vectors $\vect{H} = [\vect{h}_{1}, \ldots, \vect{h}_{M}]$ with same statistical distribution such that $M\gg N$. Then the DoF are computed by evaluating the number of linear independent vectors enclosed within this ensemble, i.e, by plotting the eigenvalues of $\vect{H} \vect{H}^{\Htran}$ in a descending order. The DoF obtained with an i.i.d. Rayleigh fading model are used as upper bounds.
In Fig.~\ref{fig:1D_DoF} we consider a 1D array of length $L_x=16\lambda$ (i.e., $N=64$ antennas). Higher dimensional arrays are illustrated in Figs.~\ref{fig:2D_DoF} and~\ref{fig:3D_DoF} respectively for a 2D array with $L_x= L_y =16\lambda$ (i.e., $N=2048$) and a 3D array with $L_x= L_y=8\lambda$ and $L_z=\lambda$ (i.e., $N=4096$). 
As seen, the theoretical expressions of the DoF computed in \eqref{DoF_1D}, \eqref{DoF_2D} and \eqref{DoF_3D} for a 1D, 2D and 3D array are validated by numerical simulations.

\begin{figure}\vspace{-0.2cm}
     \centering
     \includegraphics[width=\columnwidth]{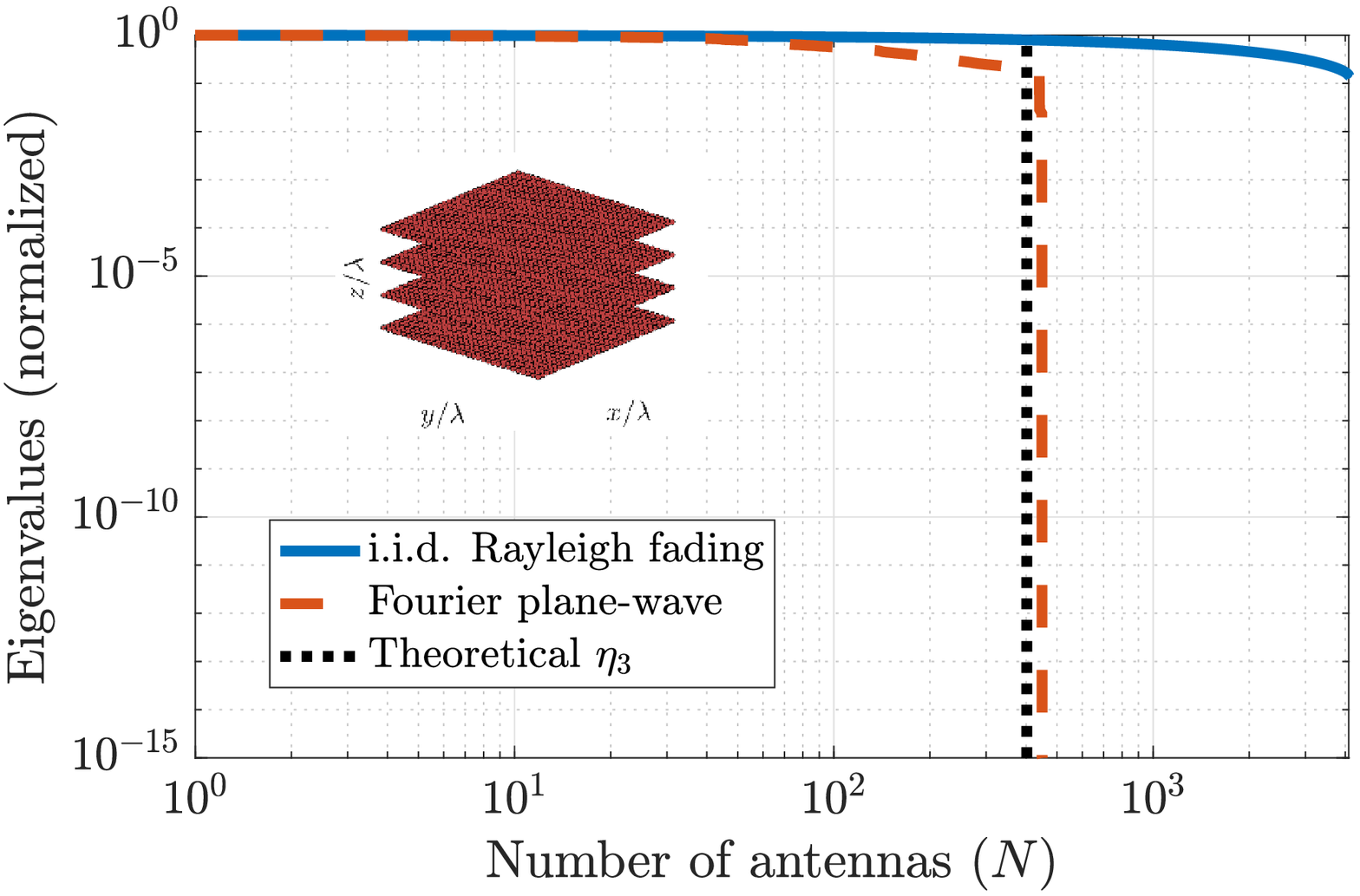}  
  \caption{Eigenvalues of $\vect{H} \vect{H}^{\Htran}$ over $\mathcal{V}_3$ in descending order.\vspace{-0.5cm}} 
   \label{fig:3D_DoF} 
\end{figure}

\section{Conclusions and Outlook}
 A random electromagnetic isotropic channel generates, over spatially-constrained apertures of rectangular symmetry, a number of spatial DoF that is proportional to its surface area, measured in units of wavelength-squared. 
The number of DoF tells us how many antennas should be deployed in a given space such that the resulting discrete array would achieve the same capacity of the continuous system. Hence, the DoF optimizes the capacity-cost tradeoff of an array by maximizing the \emph{energy efficiency} of the communication system.
Our results confirmed that the information available over a 1D segment, 2D rectangle and 3D parallelepiped can be captured by a receive discrete array that is respectively 1D with $\lambda/2$-spaced antennas, 2D with $\lambda/\sqrt{\pi}$-spaced antennas, and two parallel 2D arrays with $\lambda/\sqrt{\pi}$-spaced antennas each.

{\color{red}
The analysis focused on an isotropic scattering environment  but can be extended to the non-isotropic case. In fact, in \cite[Sec. IV.A]{pizzo2019spatial} it is shown that any non-isotropic random field can be generated by passing
an isotropic random field through a 2D linear space-invariant filter whose response (in the wavenumber domain) depends on the spectral factor $A_h(k_x,k_y,k_z)$ \cite[Lemma 1]{pizzo2019spatial}. The latter accounts for the spatial selectivity of the scattering environment. 
The isotropic spectral density $S_{h}(k_x,k_y)$ in \eqref{Fourier_planewave} is replaced by two spectral densities $S_{h}^\pm(k_x,k_y) = S_{h}(k_x,k_y) \frac{A_{h,\pm}^2(k_x,k_y)}{4\pi^2/\kappa}$ with $A_{h,\pm}(k_x,k_y) = A_h(k_x,k_y,\pm \gamma)$.
The Fourier random coefficients $H_{\ell m}^\pm$ in \eqref{Fourier_coeff_z} retain their statistical independence but have variances given by:
\begin{equation}  \label{variances}
\sigma_{\ell m, \pm}^2   = \int_{\frac{2\pi \ell}{L_x}}^{\frac{2\pi(\ell+1)}{L_x}} \int_{\frac{2\pi m}{L_y}}^{\frac{2 \pi (m+1)}{L_y}} S_{h}^\pm(k_x,k_y) \, \frac{dk_x}{2\pi} \frac{dk_y}{2\pi}.
\end{equation} 
While these variances account for the power received by the aperture from the wavenumber region indexed by the pair $(\ell,m)\in \mathcal{E}$, the support of $A_h(k_x,k_y,k_z)$ determines the cardinality of $\mathcal{E}$ from which the average number of DoF directly follows, as done for the isotropic propagation (that we recall to provide the upper limit to the number of DoF).}

\bibliographystyle{IEEEbib}
\bibliography{refs}

\begin{thebibliography}{10}

\bibitem{pizzo2019spatial}
A.~Pizzo, T.~L. Marzetta, and L.~Sanguinetti,
\newblock ``Spatially-stationary model for holographic {MIMO} small-scale
  fading,''
\newblock {\em IEEE J. Sel. Areas Commun. (to appear)}, vol. abs/1911.04853,
  2019.

\bibitem{Marzetta2018}
T.~L. {Marzetta},
\newblock ``Spatially-stationary propagating random field model for {Massive
  MIMO} small-scale fading,''
\newblock in {\em 2018 IEEE Int. Symposium Inf. Theory (ISIT)}, June 2018, pp.
  391--395.

\bibitem{Bjornson2019d}
E.~Bj\"ornson, L.~Sanguinetti, H.~Wymeersch, J.~Hoydis, and T.~L. Marzetta,
\newblock ``Massive {MIMO} is a reality---{W}hat is next? {F}ive promising
  research directions for antenna arrays,''
\newblock {\em Digital Signal Processing}, vol. 94, pp. 3--20, Nov. 2019.

\bibitem{Rusek2018}
S.~{Hu}, F.~{Rusek}, and O.~{Edfors},
\newblock ``Beyond {Massive MIMO}: The potential of data transmission with
  large intelligent surfaces,''
\newblock {\em IEEE Trans. Signal Proc.}, vol. 66, no. 10, May 2018.

\bibitem{Renzo2019a}
M.~Di Renzo, M.~Debbah, D.-T. Phan-Huy, A.~Zappone, M.-S. Alouini, C.~Yuen,
  V.~Sciancalepore, G.~C. Alexandropoulos, J.~Hoydis, H.~Gacanin, J.~de~Rosny,
  A.~Bounceu, G.~Lerosey, and M.~Fink,
\newblock ``Smart radio environments empowered by reconfigurable {AI}
  meta-surfaces: an idea whose time has come,''
\newblock {\em EURASIP Journal on Wireless Communications and Networking},
  2019.

\bibitem{Black2017}
E.~J. Black,
\newblock ``Holographic beam forming and {MIMO},''
\newblock {\em Pivotal Commware, Tech. Rep., 2017.}, 2017.

\bibitem{TseBook}
D.~Tse and P.~Viswanath,
\newblock {\em Fundamentals of Wireless Communication},
\newblock Cambridge University Press, 2005.

\bibitem{Telatar}
E.~Telatar,
\newblock ``Capacity of multi-antenna gaussian channels,''
\newblock {\em European Trans. Telecommun.}, vol. 10, no. 6, pp. 585--595,
  1999.

\bibitem{Foschini}
G.~J. Foschini and M.J. Gans,
\newblock ``On limits of wireless communications in a fading environment when
  using multiple antennas,''
\newblock {\em Wireless Personal Communications}, vol. 6, no. 3, pp. 311--335,
  Mar 1998.

\bibitem{Poon}
A.~S.~Y. Poon, R.~W. Brodersen, and D.~N.~C. Tse,
\newblock ``Degrees of freedom in multiple-antenna channels: a signal space
  approach,''
\newblock {\em IEEE Trans. Inf. Theory}, vol. 51, no. 2, Feb 2005.

\bibitem{Franceschetti}
M.~Franceschetti,
\newblock ``On {L}andau's eigenvalue theorem and information cut-sets,''
\newblock {\em IEEE Trans. Inf. Theory}, vol. 61, no. 9, Sept 2015.

\bibitem{Shannon}
C.~E. Shannon,
\newblock ``The mathematical theory of communication,''
\newblock {\em Bell System Technical Journal}, vol. 27, no. 3, pp. 379--423,
  1948.

\bibitem{GallagerBook}
Robert~G. Gallager,
\newblock {\em Principles of Digital Communication},
\newblock Cambridge University Press, 2008.

\bibitem{Landau1975}
H.~J. Landau,
\newblock ``On szeg{\"o}'s eingenvalue distribution theorem and non-hermitian
  kernels,''
\newblock {\em Journal d'Analyse Math{\'e}matique}, vol. 28, no. 1, Dec 1975.

\bibitem{Kildal2017}
P.~{Kildal}, E.~{Martini}, and S.~{Maci},
\newblock ``Degrees of freedom and maximum directivity of antennas: A bound on
  maximum directivity of nonsuperreactive antennas.,''
\newblock {\em IEEE Antennas and Propagation Magazine}, vol. 59, no. 4, Aug
  2017.

\bibitem{ChewBook}
W.~C. Chew,
\newblock {\em Waves and Fields in Inhomogenous Media},
\newblock Wiley-IEEE Press, 1995.

\bibitem{FranceschettiBook}
M.~Franceschetti,
\newblock {\em Wave Theory of Information},
\newblock Cambridge University Press, 2017.

\bibitem{Nowak}
W.G. Nowak,
\newblock ``Primitive lattice points inside an ellipse,''
\newblock {\em Czechoslovak Mathematical Journal}, vol. 55, pp. 519--530, 2005.

\bibitem{VanTreesBook}
H.~L. Van~Trees,
\newblock {\em Detection Estimation and Modulation Theory, Part I},
\newblock Wiley, 1968.

\end{thebibliography}

\end{document}